\documentclass[traditabstract]{aa}
\usepackage{natbib}
\usepackage{graphicx}
\usepackage{txfonts}
\bibpunct{(}{)}{;}{a}{}{,}
\begin{document}

\newcommand{\rmsub}[2]{#1_{\rm #2}} 
\title{The sub-Jupiter mass transiting exoplanet WASP-11b}
\author{R. G. West\inst{1} \and
A. Collier Cameron\inst{2} \and
L. Hebb\inst{2} \and
Y. C. Joshi\inst{3} \and
D. Pollacco\inst{3} \and
E. Simpson\inst{3} \and
I. Skillen\inst{4} \and
H. C. Stempels\inst{2} \and
P. J. Wheatley\inst{5} \and
D. Wilson\inst{6} \and
D. Anderson\inst{6} \and
S. Bentley\inst{6} \and
F. Bouchy\inst{7,8} \and
B. Enoch\inst{9} \and
N. Gibson\inst{2} \and
G. H\'ebrard\inst{7} \and
C. Hellier\inst{6} \and
B. Loeillet\inst{10} \and
M. Mayor\inst{11} \and
P. Maxted\inst{6} \and
I. McDonald\inst{6} \and
C. Moutou\inst{10} \and
F. Pont\inst{11} \and
D. Queloz\inst{11} \and
A. M. S. Smith\inst{2} \and
B. Smalley\inst{6} \and
R. A. Street\inst{12} \and
S. Udry\inst{11}}

\institute{Department of Physics \& Astronomy, University of Leicester,
  Leicester, LE1 7RH, UK
\and
School of Physics and Astronomy, University of St Andrews, North Haugh, St
  Andrews, Fife KY16 9SS, UK
\and
Astrophysics Research Centre, School of Mathematics \&\ Physics, Queen's
  University, University Road, Belfast, BT7 1NN, UK
\and
Isaac Newton Group of Telescopes, Apartado de Correos 321, E-38700 Santa Cruz
  de la Palma, Tenerife, Spain
\and
Department of Physics, University of Warwick, Coventry CV4 7AL, UK
\and
Astrophysics Group, Keele University, Staffordshire, ST5 5BG
\and
Institut d'Astrophysique de Paris, CNRS (UMR 7095) --  Universit\'e Pierre \&\
  Marie Curie, 98$^{bis}$ bvd. Arago, 75014 Paris, France
\and
Observatoire de Haute-Provence, 04870 St Michel l'Observatoire, France
\and
Department of Physics and Astronomy, The Open University, Milton Keynes,
MK7 6AA, UK
\and
Laboratoire d'Astrophysique de Marseille, OAMP, Universit\'e
Aix-Marseille \&
CNRS, 38 rue Fr\'ed\'eric Joliot-Curie, 13388 Marseille cedex 13, France
\and
Observatoire de Gen\`eve, Universit\'e de Gen\`eve, 51 Ch. des Maillettes,
  1290 Sauverny, Switzerland
\and
Las Cumbres Observatory, 6740 Cortona Dr. Suite 102, Santa Barbara, CA 93117,
  USA
}


\abstract{ We report the discovery of a sub-Jupiter mass exoplanet transiting
  a magnitude $\mathrm{V}=11.7$ host star 1SWASP~J030928.54+304024.7. A simultaneous
  fit to the transit photometry and radial-velocity measurements yield a
  planet mass $\rmsub{M}{p}=0.53\pm 0.07\,\mathrm{M_J}$, radius
  $\rmsub{R}{p}=0.91^{+0.06}_{-0.03}\,\mathrm{R_J}$ and an orbital period of
  $3.722465^{+0.000006}_{-0.000008}$\,days. The host star is of spectral type
  K3V, with a spectral analysis yielding an effective temperature of $4800\pm
  100\,\mathrm{K}$ and $\log g=4.45\pm 0.2$. It is amongst the smallest, least massive
  and lowest luminosity stars known to harbour a transiting exoplanet.
  WASP-11b is the third least strongly irradiated transiting exoplanet
  discovered to date, experiencing an incident flux
  $\rmsub{F}{p}=1.9\times10^{8}$\,erg\,s$^{-1}$\,cm$^{-2}$ and having an equilibrium
  temperature $\rmsub{T}{eql}=960\pm 70$\,K.  }


\date{}

\maketitle

\section{Introduction}
Observations of planets that transit their host star represent the current
best opportunity to test models of the internal structure of exoplanets and of
their formation and evolution. Since the first detection of an exoplanetary
transit signature \citep{charbonneau00, henry00} over fifty transiting
planetary systems have been identified. A number of wide-field surveys are in
progress with the goal of detecting transiting exoplanets, for example OGLE
\citep{udalski02}, XO \citep{mccullough05}, HAT \citep{bakos04}, TrES
\citep{odonovan06} and WASP \citep{pollacco06}.

The WASP project operates two identical instruments, at La Palma in the
Northern hemisphere, and at Sutherland in South Africa in the Southern
hemisphere. Each telescope has a field of view of just under 500 square degrees.
The WASP survey is sensitive to planetary transit signatures in the
light-curves of hosts in the magnitude range V\,$\sim$9--13. A detailed
description of the telescope hardware, observing strategy and pipeline data
analysis is given in \citet{pollacco06}.

In this paper we report the discovery of WASP-11b, a sub-Jupiter mass gas
giant planet in orbit about the host star 1SWASP~J030928.54+304024.7. We
present the WASP discovery photometry plus higher precision optical follow-up
and radial velocity measurements which taken together confirm the planetary
nature of WASP-11b.

\section{Observations}

\subsection{WASP photometry}

\begin{figure}
\begin{center}
\resizebox{\hsize}{!}{\includegraphics{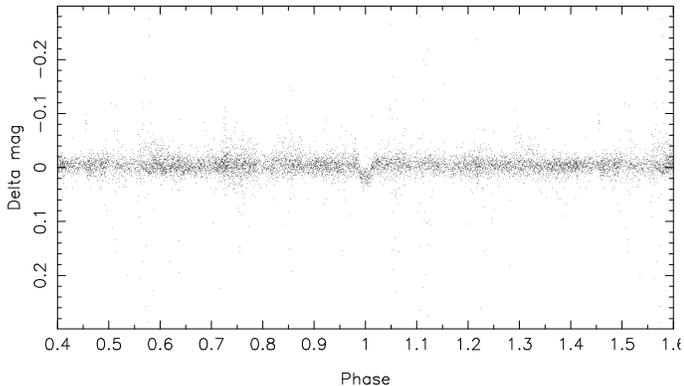}}
\caption[]{SuperWASP-N photometry of WASP-11 from the 2004, 2006 and 2007
  seasons. The data have been de-trended using the {\sc sysrem} scheme
  described in \protect\citep{colliercameron06} and are plotted here
  phase-folded on the best-fit period from the MCMC analysis (section~\ref{syspar}).}
\label{waspphot}
\end{center}
\end{figure}

The host star 1SWASP~J030928.54+304024.7 (= USNO-B1.0~1206-0003989 =
2MASS~03092855+3040249; hereafter labelled WASP-11) was
observed by SuperWASP-N during the 2004, 2006 and 2007 observing seasons,
covering the intervals 2004 July 08 to 2004 September 29, 2006 September 09 to
2007 January 20 and 2007 September 04 to 2007 December 12 respectively. The
pipeline-processed data were de-trended and searched for transits using the
methods described in \citet{colliercameron06}, yielding a detection of a
periodic transit-like signature with a period of 3.722 days. A total of ten
transits are observed in data from all three observing seasons
(Table~\ref{waspobs}; Figure~\ref{waspphot}).

\subsection{Photometric follow-up}

\begin{table}
\caption{WASP-N survey coverage of WASP-11}
\label{waspobs}
\begin{tabular}{lccccc}
Season & Camera & $N_{pts}$ & $N_{tr}$ & $T_0$ & $P$ \\
&&&&BJD-2400000.0&(days)\\
\hline
2004 & 103 & 1756 & 4 & 53240.921696 & 3.7220 \\
2006 & 144 & 2679 & 3 & 54056.140758 & 3.7223\\
2007 & 146 & 2750 & 2 & 54346.4883 & 3.7226 \\
2007 & 147 & 729 & 1 & - & - \\
\hline
\end{tabular}
\end{table}

WASP-11 was followed-up with the 2-m Liverpool telescope on La Palma as part
of the Canarian Observatories' {\it International Time Programme} for 2007-08.
We used the $2048\times 2048$ pixel EEV CCD42-40 imaging camera giving a scale
of 0.27 arcseconds/pixel in 2$\times$2 bin mode and a total field of view of
$\sim 4.6\times4.6$ arcminutes$^2$. Observations were taken during the transit
of 2008 January 14, and consist of 656 images of 10 seconds exposure in the
Sloan $z^{'}$ band. The night was non-photometric and with seeing varying from
0.9 to 2.2 arcsec during the four hour long observing run.

The images were bias subtracted and flat-field corrected with a stacked
twilight flat-field image. All the science images were also corrected for the
fringing effect. The autoguider did not work during our observations and a
maximum positional shift of 17.5 arcsec of the stars within the frame was
noticed. After aligning the images with respect to the first target image,
aperture photometry were performed around the target and comparison stars
using an aperture of 20 pixels (5\farcs 4) radius. Three bright non-variable
comparison stars were available in the target field with which to perform
differential photometry.

Further observations of WASP-11 were made with the Keele University
Observatory 60\,cm Thornton Reflector on 2008 February 09 and 13. This telescope
is equipped with a 765 $\times$ 510 pixel Santa Barbara Instrument Group
(SBIG) ST7 CCD at the f/4.5 Newtonian focus, giving a 0.68 arcsecond/pixel
resolution and a 8.63 $\times$ 5.75 arcminute field of view. Conditions were
photometric throughout both nights, although the transit of February 9 ended
at an airmass of 4 and cryogenics problems on the night of February 13 may
have led to some frosting on the CCD dewar window during the first few
exposures. Tracking errors and spurious electronic noise mean that systematic
noise is introduced into the system at an estimated level of 4 millimag with
periodicities of 2 (worm error) and 20 minutes (presently of unknown
origin). No corrections have been applied for these effects.

Altogether $(237 + 276) \times$ 30\,s observations in the R band were
obtained. After applying corrections for bias, dark current and flat fielding
in the usual way, aperture photometry on WASP-11 and the comparison star
USNO-B1.0 1207-0040657 were performed using the commercial software AIP4Win
\citep{berry05}. The resulting lightcurves from both Liverpool Telescope and
Keele 60\,cm observations (Figure~\ref{mcmcfit} top panel) confirm the presence
of a transit.

\subsection{Radial velocity follow-up}

Initial spectroscopic observations were obtained using the FIbre-fed Echelle
Spectrograph (FIES) mounted on the 2.5-m Nordic Optical Telescope. A total of
five radial velocity points were obtained during 2007 December 27--31 and
2008 January 25. WASP-11 was observed with an exposure time of 1800s giving a
signal-to-noise ratio  of around 70--80 at 5500{\AA}. FIES was used in
medium resolution mode with R=46000 with simultaneous ThAr calibration.  We
used the bespoke data reduction package FIEStool to extract the
spectra and a specially developed IDL line-fitting code to obtain radial
velocities with a precision of 20--25m\,s$^{-1}$. 

Radial velocity measurements of WASP-11 were also made with the Observatoire
de Haute-Provence's 1.93\,m telescope and the SOPHIE spectrograph
\citep{bouchy}, over the 8 nights 2008 February 11 -- 15; a total of 7 usable
spectra were acquired. SOPHIE is an environmentally stabilized spectrograph
designed to give long-term stability at the level of a few m\,s$^{-1}$. We
used the instrument in its medium resolution mode, acquiring simultaneous star
and sky spectra through separate fibres with a resolution of R=48000.
Thorium-Argon calibration images were taken at the start and end of each
night, and at 2- to 3-hourly intervals throughout the night.  The
radial-velocity drift never exceeded 2--3 m\,s$^{-1}$, even on a
night-to-night basis.

\begin{table}
\caption{Radial velocity measurements of WASP-11}
\label{rvobs}
\begin{tabular}{ccccc}
BJD & RV & $\sigma_{RV}$ & $v_{span}$ & Inst \\
(UT) & (km\,s$^{-1}$) & (km\,s$^{-1}$) & \\
\hline
2454462.395   & 4.8689 & 0.0185 & & NOT \\
2454463.456   & 4.8725 & 0.0203 & & NOT \\
2454465.404   & 4.9208 & 0.0258 & & NOT \\
2454466.440   & 4.8262 & 0.0244 & & NOT \\
2454466.443   & 4.9486 & 0.0246 & & NOT \\
2454491.424   & 4.9339 & 0.0220 & & NOT \\
2454508.3700 & 4.8910 & 0.0103  & 0.011 & SOPHIE \\
2454509.3534 & 5.0104 & 0.0084  & 0.000 & SOPHIE \\
2454510.3813 & 4.8989 & 0.0120  & 0.025 & SOPHIE \\
2454511.3092 & 4.8515 & 0.0076  & -0.002 & SOPHIE \\
2454511.3800 & 4.8330 & 0.0106  & -0.002 & SOPHIE \\
2454511.4206 & 4.8235 & 0.0143  & -0.035 & SOPHIE \\
2454512.3848 & 4.9482 & 0.0096  & 0.022 & SOPHIE \\
\hline
\end{tabular}
\end{table}

Conditions during the SOPHIE observing run were photometric throughout, though
all nights were affected by strong moonlight. Integrations of 1080\,s yielded a
peak signal-to-noise per resolution element of around $\sim$30--40. The
spectra were cross-correlated against a K5V template provided by the SOPHIE
control and reduction software.

\begin{figure}
\begin{center}
\resizebox{\hsize}{!}{\includegraphics{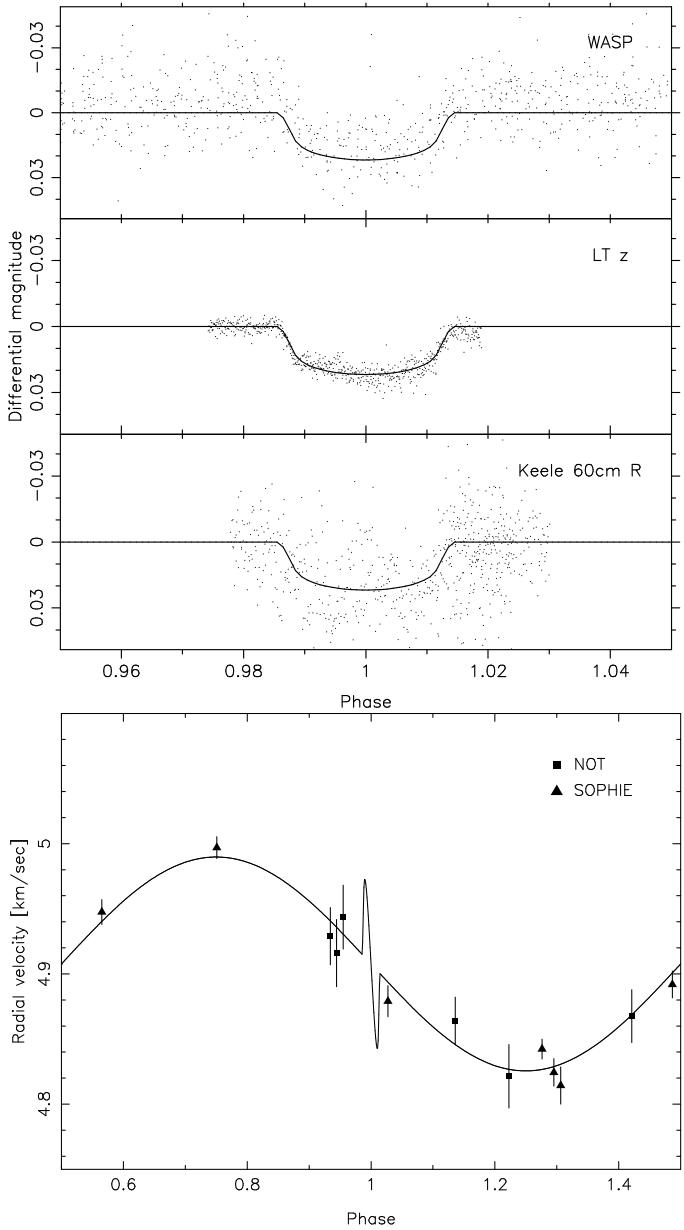}}
\vspace{5mm}
\resizebox{\hsize}{!}{\includegraphics{rvplot.ps}}
\caption[]{The best-fit model from the simultaneous MCMC fit to the available
  photometry (top panel) and radial velocity data (lower panel). The fitted
  zero-point offset between the NOT and SOPHIE radial-velocity measurements
  ($5.4\pm 0.4$m\,s$^{-1}$) is removed in this plot.}
\label{mcmcfit}
\end{center}
\end{figure}

In all SOPHIE spectra the cross-correlation functions (CCF) were contaminated by the
strong moonlight. We corrected them by using the CCF from the background
light's spectrum (mostly the Moon) in the sky fibre. We then scaled both CCFs
using the difference of efficiency between the two fibres.  Finally we
subtracted the corresponding CCF of the background light from the star fibre,
and fitted the resulting function by a Gaussian. The parameters obtained allow
us to compute the photon-noise uncertainty of the corrected radial velocity
measurement ($\rmsub{\sigma}{RV}$), using the relation
$$
\rmsub{\sigma}{RV} = 3.4\sqrt{(FWHM)}/(S/N\times Contrast)
$$
Overall our SOPHIE RV measurements have an average photon-noise uncertainty
of 10.3\,m\,s$^{-1}$. The measured barycentric radial
velocity (Table~\ref{rvobs}, Figure~\ref{mcmcfit} lower panel) show
a sinusoidal variation of half-amplitude $\sim 90$\,m\,s$^{-1}$ about a centre-of-mass
RV of $\sim 4.9$\,km\,s$^{-1}$, consistent with the presence a companion of
planetary mass. The period and ephemeris of the RV variation are consistent
with those of found by the transit search.

\begin{figure}
\begin{center}
\resizebox{\hsize}{!}{\includegraphics{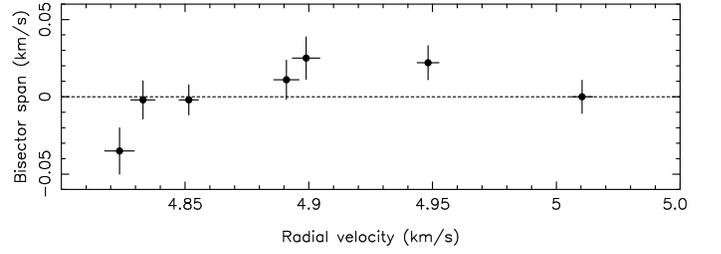}}
\caption[]{The line bi-sector against velocity for WASP-11, showing no
  evidence of correlation.}
\label{rvbs}
\end{center}
\end{figure}

An analysis of the line-bisector spans shows no significant correlation with
radial velocity (Figure~\ref{rvbs}), as would be expected if the observed
radial velocity variations were due to a diluted eclipsing binary or
chromospheric activity \citep{queloz}.

\section{System parameters}\label{syspar}
\subsection{Stellar parameters}
In order to perform a detailed spectroscopic analysis of the stellar
atmospheric properties of WASP-11, we merged the available FIES spectra into
one high-quality spectrum, carefully removing any radial velocity signature
during the process. This merged spectrum was then continuum-normalized with a
very low order polynomial to retain the shape of the broadest spectral
features. The total signal-to-noise of the combined spectrum was around 200
per resolution element. We were not able to include the SOPHIE spectra in this
analysis, because these spectra were obtained with the HE (high-efficiency)
mode which is known to suffer from problems with removal of the blaze
function.

\begin{table}
\caption{System parameters of WASP-11 derived from a simultaneous MCMC
  analysis of the available photometric and radial-velocity measurements.
  Quoted uncertainties define the $1\sigma$ confidence intervals.}
\label{mcmcresults}
\begin{tabular}{lcc}
\hline
Transit epoch (HJD), $T_0$ & $2454473.05586\pm 0.0002$ &\\
Orbital period, $P$ & $3.722465^{+0.000006}_{-0.000008}$ & days \\
$(\rmsub{R}{p}/R_\star)^2$ & $0.0162^{+0.0003}_{-0.0002}$ & \\
Transit duration & $2.556^{+0.029}_{-0.007}$ & hours \\
Impact parameter, $b$ & $0.054^{+0.168}_{-0.050}$ & $\mathrm{R_*}$ \\

Reflex velocity, $K_1$ & $0.0821\pm 0.0074$ & km\,s$^{-1}$ \\
Centre-of-mass velocity, $\gamma$ & $4.9077\pm 0.0015$ & km\,s$^{-1}$ \\

Orbital eccentricity, $e$ & $\equiv 0.0$ & \\
Orbital inclination, $i$ & $89.8^{+0.2}_{-0.8}$ & deg \\
Orbital separation, $a$ & $0.043\pm 0.002$ & AU \\

Stellar mass, $M_\star$ & $0.77^{+0.10}_{-0.08}$ & $\mathrm{M_{\odot}}$ \\
Stellar radius, $R_\star$ & $0.74^{+0.04}_{-0.03}$ & $\mathrm{R_{\odot}}$ \\
  
Planet radius, $\rmsub{R}{p}$ & $0.91^{+0.06}_{-0.03}$ & $\mathrm{R_J}$ \\
Planet mass, $\rmsub{M}{p}$ & $0.53\pm 0.07$ & $\mathrm{M_J}$ \\

$\log \rmsub{g}{p}$ (cgs) & $3.16^{+0.04}_{-0.05}$ & \\
Planet density, $\rmsub{\rho}{p}$ & $0.69^{+0.07}_{-0.11}$ & $\rho_J$ \\
Planet $\rmsub{T}{eql}$ (A=0; f=1) & $960\pm 70$ & K \\

\hline
\end{tabular}
\end{table}

\begin{figure*}
\begin{center}
\includegraphics[angle=90,width=17cm]{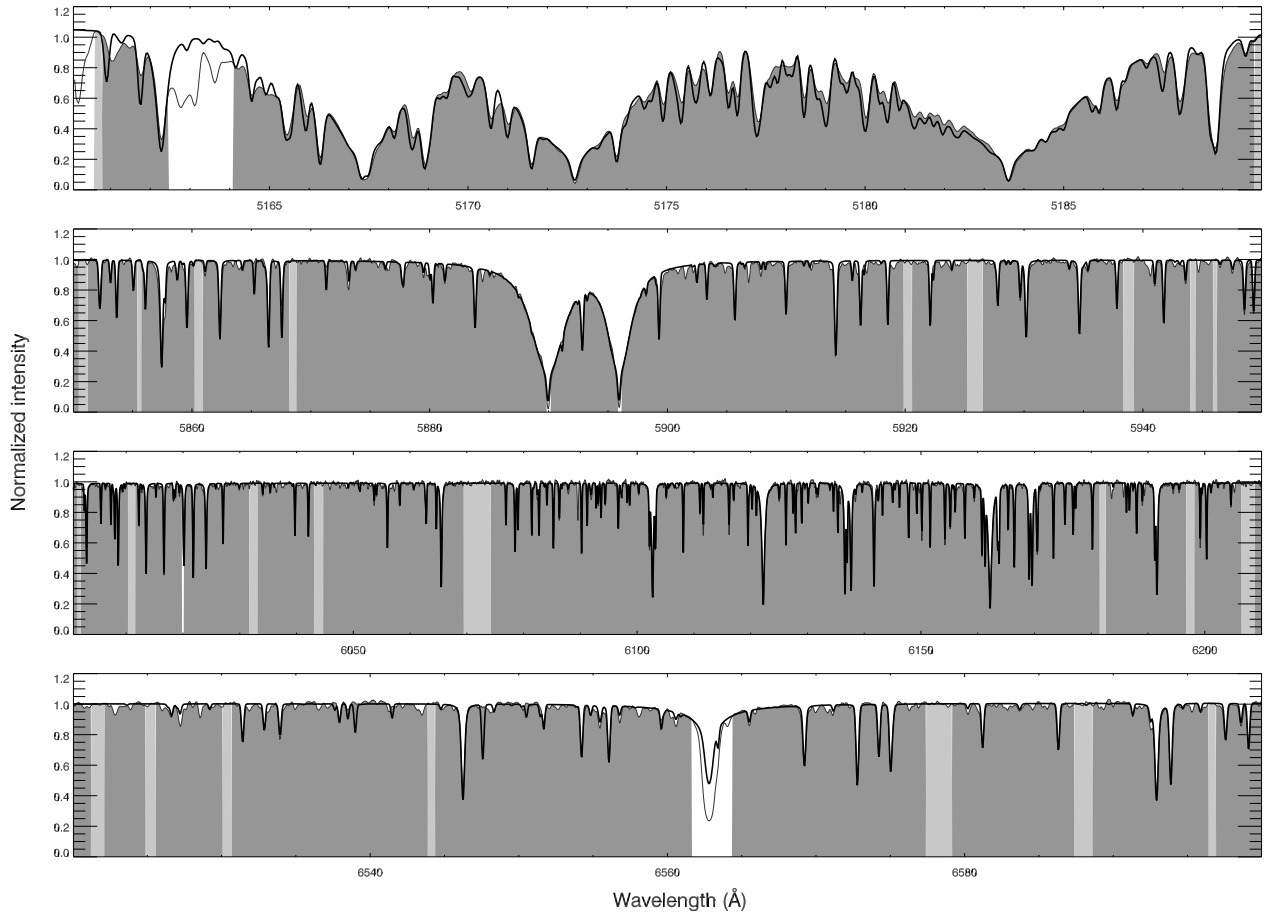}
\caption[]{A comparison between the observed FIES spectrum of WASP-11 and
  the calculated spectrum obtained from spectral synthesis with SME. The white
  regions are excluded from the spectral analysis, mainly because of the
  presence of telluric absorption. Light shaded regions were used to determine
  the continuum level, and the remaining dark shaded regions to determine the
  stellar atmospheric parameters.}
\label{specfit}
\end{center}
\end{figure*}
For our analysis we followed the same procedure as for the spectroscopic
characterization of WASP-1 \citep{stempels} and WASP-3 \citep{pollacco08}. We
used the package Spectroscopy Made Easy \citep[SME,][]{valenti}, which combines
spectral synthesis with multidimensional $\chi^2$ minimization to determine
which atmospheric parameters best reproduce the observed spectrum of WASP-11
(effective temperature $T_{\rm eff}$, surface gravity $\log g$, metallicity
[M/H], projected radial velocity $v \sin i$, systemic radial velocity $v_{\rm
  rad}$, microturbulence $v_{\rm mic}$ and the macroturbulence $v_{\rm mac}$).
For a more detailed description of the spectral synthesis and our assumptions
we refer to \citet{stempels}.

The four spectral regions we used in our analysis are (1) 5160--5190{\AA},
covering the gravity-sensitive Mg b triplet (2) 5850--5950{\AA}, with the
temperature and gravity-sensitive Na {\sc i} D doublet (3) 6000-6210{\AA},
containing a wealth of different metal lines, providing leverage on the
metallicity, and (4) 6520--6600{\AA}, covering the strongly
temperature-sensitive H-alpha line. A comparison between the observed FIES
spectrum and the synthetic spectrum is shown in Figure~\ref{specfit}. The
spectral analysis yields an effective temperature $T_{\rm eff}=4800\pm
100$\,K, $\log g=4.45\pm 0.2$, $[M/H]=0.0\pm 0.2$ and $v\sin
i<6.0$\,km\,s$^{-1}$. These parameters correspond to spectral type of K3V. A
close examination of the region around the Li {\sc i} 6708 shows no evidence
of such a feature, suggesting that the lithium abundance is very low.

\subsection{Planet parameters}
To determine the planetary and orbital parameters the SOPHIE and NOT FIES
radial velocity measurements were combined with the photometry from WASP and
the Liverpool Telescope in a simultaneous fit using the Markov Chain Monte
Carlo (MCMC) technique. The details of this process are described in
\citet{pollacco08}. An initial fit showed that the orbital eccentricity
($e=0.086^{+0.070}_{-0.062}$) was poorly constrained by the available data and
nearly consistent with zero. We therefore fixed the eccentricity parameter at
zero in a further fits. Figure~\ref{mcmcfit} shows the best-fitting models.
The best-fit parameters (Table~\ref{mcmcresults}) show WASP-11b to have a mass
$M=0.53\pm 0.07\,\mathrm{M_J}$ and a radius of
$R=0.91^{+0.06}_{-0.03}\,\mathrm{R_J}$.


\section{Discussion}

\begin{figure}
\begin{center}
\resizebox{\hsize}{!}{\includegraphics{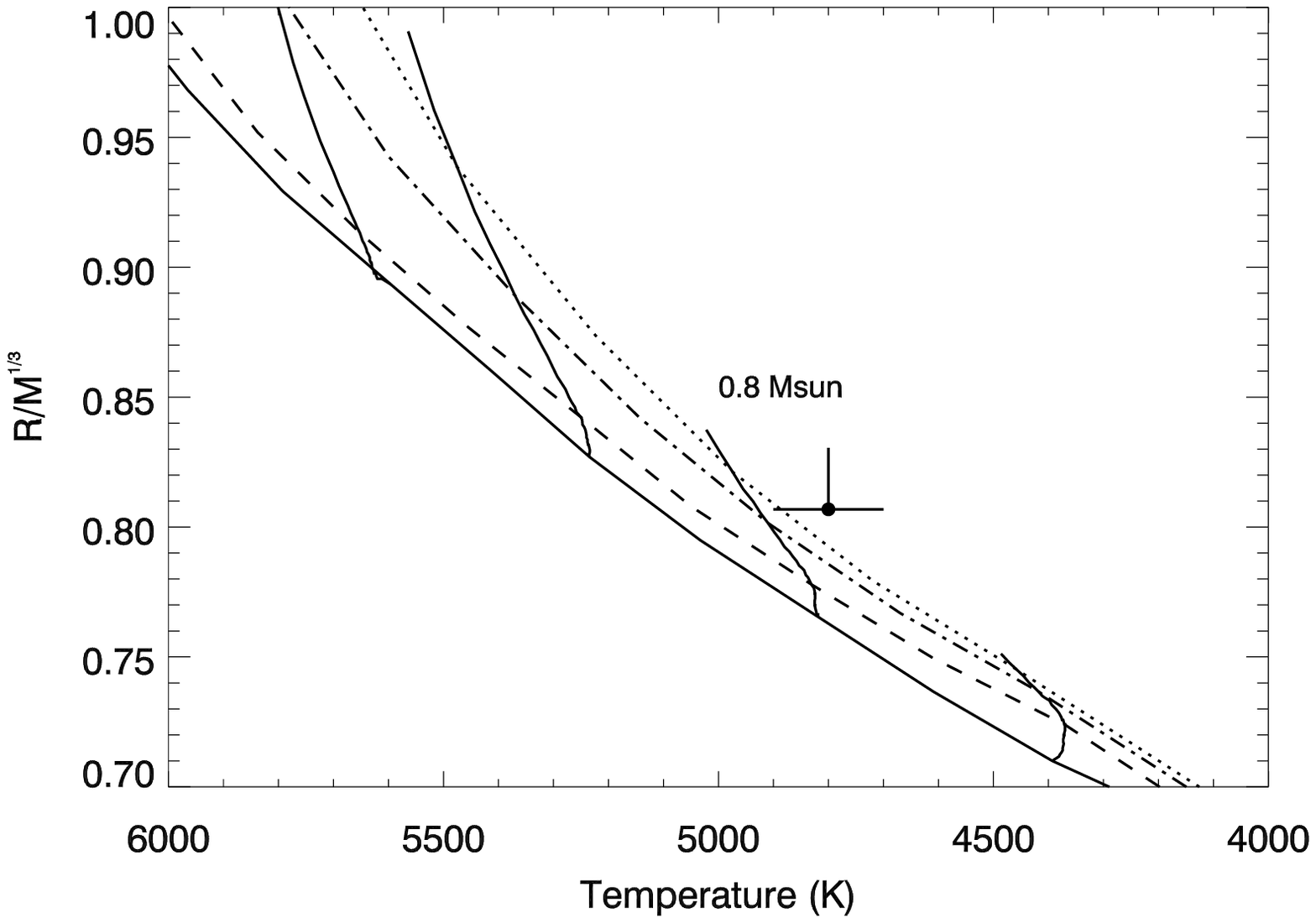}}
\resizebox{\hsize}{!}{\includegraphics{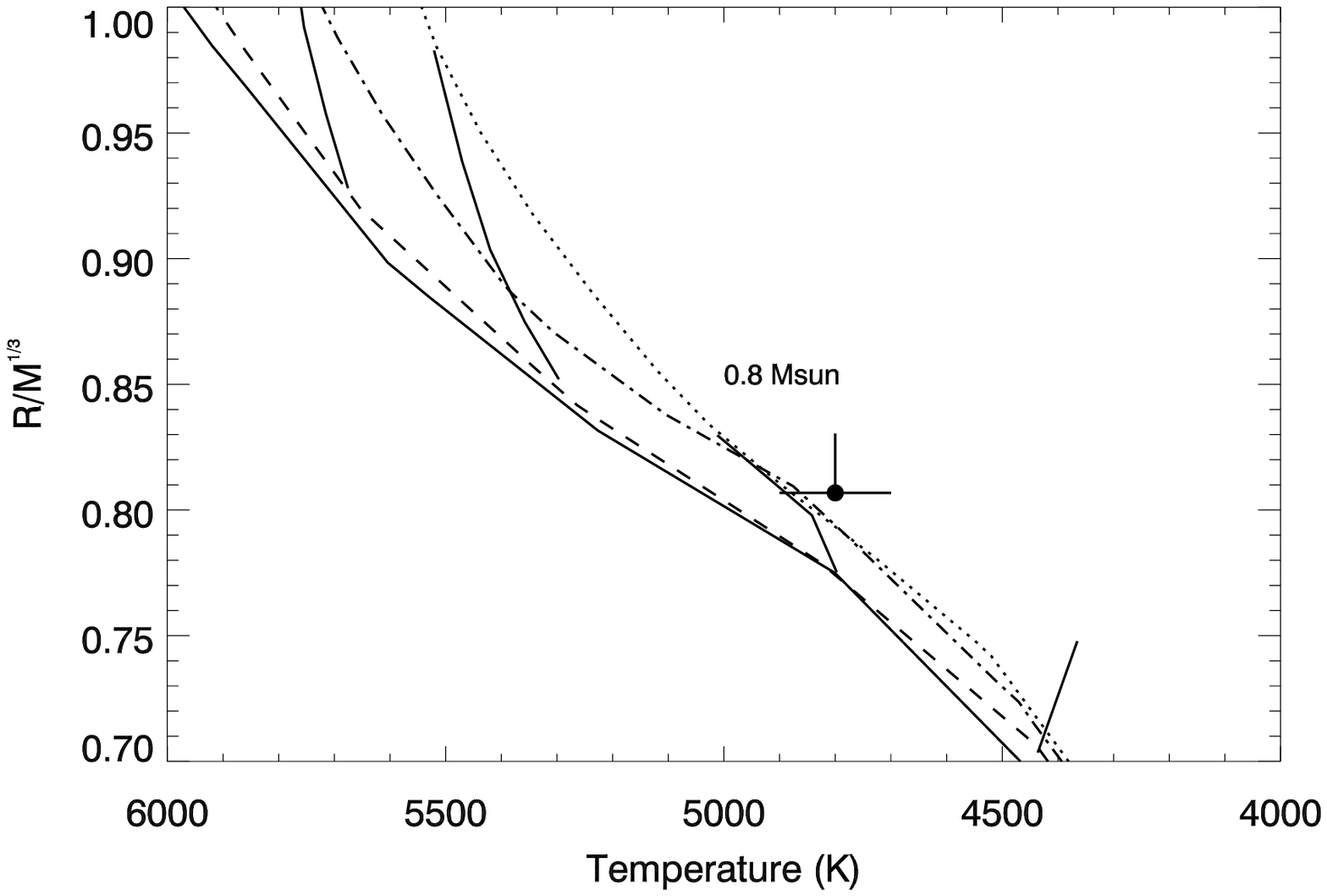}}
\caption[]{The position of WASP-11 in the $R/M^{1/3}-\rmsub{T}{eff}$
  plane. Evolutionary tracks for a solar metallicity star from
  \citet{baraffe98} (upper panel) and \citet{girardi00} (lower panel) are
  plotted along with isochrones for ages 10\,Myr (solid), 1\,Gyr (dashed), 5\,Gyr
  (dot-dashed), 10\,Gyr (dotted). Evolutionary mass tracks are shown for 0.7,
  0.8, 0.9 and 1.0\,M$_\odot$.}
\label{starevol}
\end{center}
\end{figure}

The system parameters derived here place WASP-11b towards the lower end of the
mass range of known transiting planets, falling approximately mid-way between
the masses of Jupiter and Saturn. The host star WASP-11 is also amongst the
smallest and lowest luminosity stars known to host a transiting planet,
however it is relatively nearby and thus quite bright ($\mathrm{V}=11.7$). WASP-11b is
irradiated by a stellar flux $\rmsub{F}{p}=1.9\times
10^8$\,erg\,cm$^{-2}$\,s$^{-1}$ at the sub-stellar point making it the third
least heavily irradiated transiting planet after GJ436b and HD17156b. We
compute an equilibrium temperature for WASP-11b of $\rmsub{T}{eql}(A=0;
f=1)=960\pm70$\,K, which makes it more typical of the bulk of known exoplanets
than of the ``hot Jupiter'' class most commonly found by the transit method.

Theoretical models of the atmospheres of hot giant exoplanets
\citep{fortney06, burrows07} have shown that heavy irradiation can lead to the
development of a temperature inversion and a hot stratosphere.  This is due to
the absorption of stellar flux by an atmospheric absorber, possibly TiO and
VO. In both sets of models the magnitude of the incident stellar flux is the
key controlling variable determining whether a given extra-solar giant planet
(EGP) will possess a hot stratosphere. Recent observations by
\citet{machalek08} of secondary transits of XO-1b using the {\it Spitzer Space
  Telescope} suggest the presence of a temperature inversion in the atmosphere
of that exoplanet.  On the other hand analogous observations of HD189733b
\citep{charbonneau08} show no evidence for an inversion, despite the
irradiating fluxes of XO-1b and HD189733b being almost identical
($\rmsub{F}{p}=0.49\times 10^9$ and $\rmsub{F}{p}=0.47\times
10^9$\,erg\,cm$^{-2}$\,s$^{-1}$ respectively). This strongly suggests that the
incident stellar flux is not the sole controlling parameter determining the
presence of the inversion, a likelihood which the authors of the atmosphere
models readily point out themselves. Further observations of planets
particularly in the low-irradiation regime are required to help parameterise
the thermal inversion. WASP-11b is amongst the nearest and brightest
low-irradiation EGPs making it a good candidate for such studies. Moreover we
note that the orbital eccentricity of WASP-11b is much lower than the other
two bright low-irradiation transiting exoplanets, GJ436b and HD17156b
($e=0.15$ and $e=0.67$ respectively). As a consequence the secular variation
in irradiation around the orbit will be correspondingly lower in WASP-11b,
removing a potentially complicating factor when comparing follow-up
observations with predictions from atmospheric models developed assuming
steady-state irradiation.

To estimate the age of the WASP-11 we compared the observed stellar density
and temperature against the evolutionary models of low- and intermediate-mass
stars of \citet{girardi00} and \citet{baraffe98}. In Figure~\ref{starevol} we
plot the position of WASP-11 in the $R/M^{1/3}$ versus $\rmsub{T}{eff}$ plane
atop isochrones of different ages from the two models. For such a cool star,
the isochrones are closely spaced in this parameter plane due to the slow
post-main-sequence evolution of late-type stars. The sets of isochrones from
the two models overlap in this regime, and both models suggest the same mass
and age for the host star. WASP-11 falls above the 10\,Gyr isochrone for both
models, though it is consistent with this age within the errors. The very low
lithium abundance also points toward WASP-11 being $\ga 1$--2\,Gyr old
\citep{sestito05}. We investigated using gyrochronology to age the host star,
following \citet{barnes07}, however we were unable to measure a definite
rotational period. No rotation modulation was detected in the lightcurve to an
amplitude limit of a few milli-magnitudes. The spectral analysis furnishes
only an upper-limit to $v\sin i$, so no rotational period can be determined in
that way. Taken together these factors are all consistent with WASP-11 being
an old star, older than maybe 1\,Gyr, however it is not possible to be more
definite than that with the available data.

\begin{figure}
\begin{center}
\resizebox{\hsize}{!}{\includegraphics{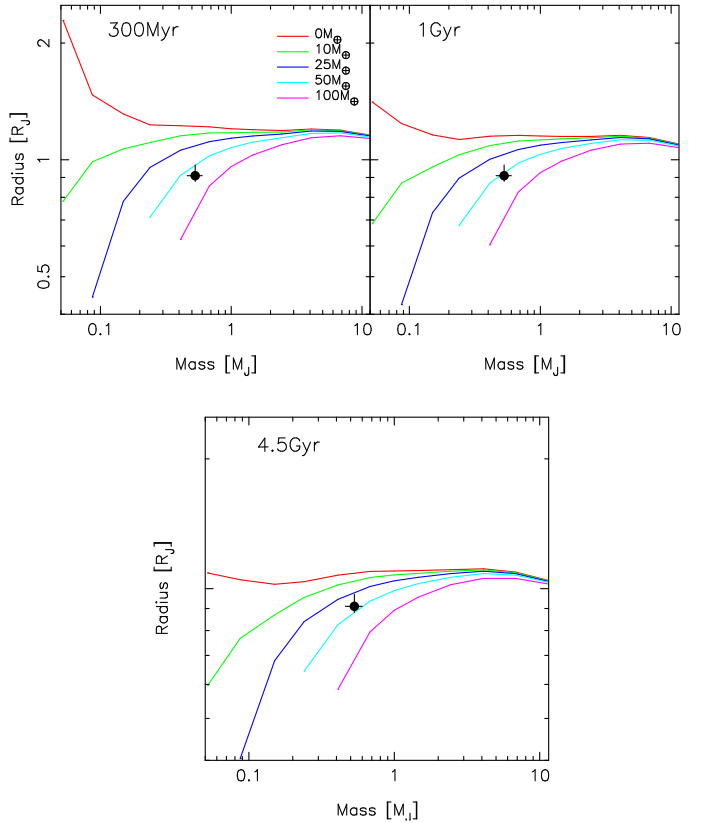}}
\caption[]{Planetary mass-radius relations as a function of core mass and
  system age, interpolated from the models of \protect\citet{fortney07}.}
\label{coremass}
\end{center}
\end{figure}

\citet{fortney07} present models of the evolution of planetary radius over a
range of planetary masses and orbital distances, and under the assumption of
the presence of a dense core of various masses up to 100\,M$_\oplus$.  To
compare our results with the Fortney et al. models we plotted the modelled
mass-radius relation as a function of core mass in Figure~\ref{coremass}. To
account for the lower-than-Solar luminosity of the host star WASP-11 we
calculated the orbital distance $a_\odot=a(M_\star/M_\odot)^{-3.5/2}$ at which
a planet in orbit about the Sun would receive the same incident stellar flux
as WASP-11b does from its host. We then interpolated the models of Fortney et
al. to this effective orbital distance ($a_\odot=0.068$ for WASP-11b). As the
age of the WASP-11 system is poorly constrained we compare our results with the
modelled mass-radius relation at 300\,Myr, 1\,Gyr and 4.5\,Gyr. We find that the
radius of WASP-11b is consistent with the presence of a dense core with a mass
in the range $\rmsub{M}{core}\sim$ 42--77\,M$_{\oplus}$ for a system age of
300\,Myr, $\rmsub{M}{core}\sim$33--67\,M$_{\oplus}$ at 1\,Gyr, and
$\rmsub{M}{core}\sim$22--56\,M$_{\oplus}$ at 4.5\,Gyr.

\begin{acknowledgements}
The WASP Consortium consists of astronomers primarily from the Queen's
University Belfast, Keele, Leicester, The Open University, and St Andrews, the
Isaac Newton Group (La Palma), the Instituto de Astrof{\'i}sica de Canarias
(Tenerife) and the South African Astronomical Observatory. The SuperWASP-N and
WASP-S Cameras were constructed and operated with funds made available from
Consortium Universities and the UK's Science and Technology Facilities
Council.  SOPHIE observations have been funded by the Optical Infrared
Coordination network (OPTICON), a major international collaboration supported
by the Research Infrastructures Programme of the European Commission's Sixth
Framework Programme. FIES observations were made with the Nordic Optical
Telescope, operated on the island of La Palma jointly by Denmark, Finland,
Iceland, Norway, and Sweden, in the Spanish Observatorio del Roque de los
Muchachos of the Instituto de Astrofisica de Canarias. We extend our thanks to
the Director and staff of the Isaac Newton Group of Telescopes for their
support of SuperWASP-N operations, and the Director and staff of the
Observatoire de Haute-Provence for their support of the SOPHIE spectrograph.
The Liverpool Telescope is operated on the island of La Palma by Liverpool
John Moores University in the Spanish Observatorio del Roque de los Muchachos
of the Instituto de Astrofisica de Canarias with financial support from the UK
Science and Technology Facilities Council.
\end{acknowledgements}


\end{document}